\documentclass[psamsfonts,reqno]{amsart}

\usepackage{amsmath}
\usepackage{amssymb}
\usepackage{color}
\usepackage{mathrsfs}
\usepackage{bbm}

\usepackage{graphicx}
\usepackage{hyperref}
\usepackage{apacite}          

\usepackage{tabularx}
\usepackage{booktabs}
\usepackage{natbib}
\usepackage{setspace}

\usepackage{float}
\usepackage{subfig}
\usepackage{graphicx}

\newtheorem{theorem}{Theorem}

  \title[Pricing Path-dependent Options under Stochastic Volatility]{Pricing 
  Path-dependent Options under Stochastic Volatility via Mellin Transform}
  
  \author[J. Cao]{Jiling Cao}
  \address{Department of Mathematical Sciences, School of Engineering, Computer and 
  Mathematical Sciences, Auckland University of Technology, Private Bag 92006, Auckland 
  1142, New Zealand}
  \email{jiling.cao@aut.ac.nz}
	
  \author[J.-H. Kim]{Jeong-Hoon Kim}
  \address{Department of Mathematics, Yonsei University, Seoul 03722, Republic of Korea}
  \email{jhkim96@yonsei.ac.kr}
	
  \author[X. Li]{Xi Li}
  \address{Department of Mathematical Sciences, School of Engineering, Computer and 
  Mathematical Sciences, Auckland University of Technology, Private Bag 92006, Auckland 
  1142, New Zealand}
  \email{xi.li@aut.ac.nz}
	
  \author[W. Zhang]{Wenjun Zhang}
  \address{Department of Mathematical Sciences, School of Engineering, Computer and 
  Mathematical Sciences, Auckland University of Technology, Private Bag 92006, Auckland 
  1142, New Zealand}
  \email{wenjun.zhang@aut.ac.nz}
  
  \date{}
  
  \thanks{\hspace{-1.66em} 2020 \emph{Mathematics Subject Classification.}
  Primary 91G20; Secondary 41A60, 44A99, 91G60.}
  \thanks{\noindent \emph{Keywords and phrases}. Asymptotic approximation, barrier, 
  down-and-out, floating strike, lookback, Mellin transform, stochastic volatility.}

\begin{document}
	
  \maketitle

  \begin{abstract}
  In this paper, we derive closed-form formulas of first-order approximation for 
  down-and-out barrier and floating strike lookback put option prices under a stochastic 
  volatility model, by using an asymptotic approach. To find the explicit closed-form 
  formulas for the zero-order term and the first-order correction term, we use Mellin 
  transform. We also conduct a sensitivity analysis on these formulas, and compare the 
  option prices calculated by them with those generated by Monte-Carlo simulation.   
  \end{abstract}
	
  \section{Introduction}
  
  A standard option gives its owner the right to buy (or sell) some underlying asset 
  in the future for a fixed price. Call options confer 
  the right to buy the asset, while put options confer the right to sell the asset. 
  Path-dependent options represent extensions of this concept. For example, a lookback 
  call option confers the right to buy an asset at its minimum price over some time 
  period. A barrier option resembles a standard option except that the payoff also 
  depends on whether or not the asset price crosses a certain barrier level during the
  option’s life. Lookback and barrier options are two of the most popular types 
  of path-dependent options.
  
  Following the lead set by \cite{Bla1973} and assuming that the underlying asset 
  price follows geometric Brownian motion with constant volatility, \cite{Mer1973} 
  derived a closed-form pricing formula for down-and-out call options. 
  \cite{Reiner91} extended Merton’s result to other types of barrier options. 
  \cite{Gold79a}, \cite{Gold79b} and \cite{Con91} provided closed-form pricing 
  formulas for lookback options. For a good summary for research on path-dependent 
  options under Black-Scholes framework, refer to \cite{Clew94}. As we know, the 
  assumption that an asset price process follows geometric Brownian 
  motion with constant volatility does not capture the empirical observations, 
  due to the volatility smile effect. So, it is desirable to overcome this 
  drawback. There are different ways of extending the Black-Scholes model to 
  incorporate the ``smile" feature: one way is to consider ``local volatility" and 
  the other is to consider ``stochastic volatility".
  
  One of popular local volatility models was the constant elasticity of variance (CEV)
  model introduced by \cite{Cox1975}, where a closed-form pricing formula for European
  call options was presented. \cite{Davy01} derived solutions for barrier and lookback
  option prices under the CEV process in closed form, and demonstrated that barrier 
  and lookback option prices and hedge ratios under the CEV process can deviate 
  dramatically from the lognormal values. In \cite{boyle99}, the pricing of certain 
  path-dependent options was re-examined when the underlying asset follows the CEV 
  diffusion process, by approximating the CEV process using a trinomial method.
  
  In general, the pricing problems of path-dependent options do not have analytic 
  solutions under stochastic volatility. \cite{Chia12} considered the problem of 
  numerically evaluating barrier option prices when the dynamics of the underlying 
  are driven by Heston stochastic volatility model and developed a method of lines 
  approach to evaluate the price as well as the delta and gamma of the option. 
  \cite{Park13} investigates a semi-analytic pricing method for lookback options 
  in a general stochastic volatility framework. The resultant formula is well 
  connected to the Black–Scholes price that is the first term of a series expansion, 
  which makes computing the option prices relatively efficient. Further, a 
  convergence condition for the expansion is provided with an error bound.
  \cite{leung13} and \cite{wirtu17} derived an analytic pricing formula for 
  floating strike lookback options under the Heston model by means of the homotopy 
  analysis method. The price is given by an infinite series whose value can be 
  determined once an initial term is given well.
 
  In addition, \cite{Kato13} derived a new semi closed-form approximation formula 
  for pricing an up-and-out barrier option under a certain type of stochastic 
  volatility model including SABR model. In a more recent paper by \cite{Fun2018}, 
  a unified approximation scheme was proposed for a single barrier option under local 
  volatility models, stochastic volatility models, and their combinations. The basic 
  idea of their approximation is to mimic a target underlying asset process by a 
  polynomial of the Wiener process. They then translated the problem of solving the 
  first hit probability of the asset price into the problem of solving that of a Wiener 
  process whose distribution of the passage time is known. Finally, utilizing Girsanov’s 
  theorem and the reflection principle, they showed that single barrier option prices 
  can be approximated in a closed-form.
  
  The main contribution of this paper is to derive new closed-form approximation 
  formulas for pricing down-and-out put barrier options and floating strike lookback
  put options under a certain type of stochastic volatility model, which is similar to
  the one in \cite{Kato13}. To achieve our goal, we apply the asymptotic approach 
  discussed in \cite{fou11} and Mellin transform. Mellin transform techniques 
  were used by \cite{Panini04} to derive integral equation representations for the 
  price of European and American basket put options. Similarly, \cite{Yoon14} 
  applied Mellin transform to derive a closed form solution of the option price 
  with respect to a European call option and a European put option with Hull-White 
  stochastic interest rate. Moreover, \cite{Kim18} derived a closed-form formula of 
  a second-order approximation for a European corrected option price under stochastic 
  elasticity of variance (SEV) model.
  
  The rest of the paper is organized as follows. Section \ref{sec:modelling} discusses 
  the model framework and the features of down-and-out and floating strike lookback
  put options. In Section \ref{sec:expansion}, we give detailed discussions on an 
  asymptotic approach which is used to derive approximations to the risk-netural 
  values of these types of options. In Section \ref{sec:barrier}, we apply Mellin 
  transform to derive a closed-form formula of the first-order approximation for 
  down-and-out barrier put options. In Section \ref{sec:lookback}, we apply Mellin 
  transform to derive a closed-form formula of the first-order approximation for 
  floating strike barrier put options. Section \ref{sec:numerical} presents 
  sensitivity and comparison analysis, and demonstrate that the results given by 
  these closed-form formulas match well with those generated by Monte-Carlo 
  simulation. Section \ref{sec:conclusion} gives a brief summary. Details on 
  Mellin transform and derivation of the closed-form formulas in Sections 
  \ref{sec:barrier} and \ref{sec:lookback} are provided in Appendices A and B, 
  respectively.

  \section{Basic Model Set-up and Path-dependent Options} \label{sec:modelling}
  
  \subsection{Stochastic volatility model} 
  Let $\{S_t:t\geq 0 \}$ denote the price process of a risky asset on some filtered 
  probability space $\left(\Omega, {\mathscr F}, ({\mathscr F}_t)_{t\ge 0}, {\mathbb P}
  \right)$, where $\mathbb P$ is the physical probability measure. In this paper, we 
  assume that $\{S_t:t\geq 0 \}$ evolves according to the following system of stochastic 
  differential equations: 
  \begin{eqnarray}\label{eqn:modeling_frame}
  dS_t&=&\mu S_t dt+f\left(Y_t\right)S_t  dW^s_t, \nonumber \\ 	
  dY_t&=&\alpha\left(m-Y_t\right)dt+\beta \left(\rho dW^s_t +
  \sqrt{1-\rho^2} dW_t^y\right), 
  \end{eqnarray}
  where $\mu$, $\alpha>0$, $\beta >0$ and $m$ are constants, $f$ is a function having 
  non-zero values and specifying the dependence on the hidden process $\{Y_t:t\geq 0 \}$. 
  The processes $\left\{W_t^s:t\geq 0 \right\}$ and $\left\{W_t^y: t\geq 0 \right\}$ 
  are independent standard Brownian motions. The constant correlation coefficient 
  $\rho$ with $-1 < \rho <1$ captures the leverage effect. Here, $\mu$ is the drift 
  rate. The mean-reversion process $\{Y_t:t\geq 0 \}$ given in 
  Eq. (\ref{eqn:modeling_frame}) is characterized by its typical time to obtain back 
  to the mean level $m$ of its long-run distribution. The parameter $\alpha$ determines 
  the speed of mean-reversion and $\beta$ controls the volatility of $\{Y_t:t\geq 0 \}$. 
  In the sequel, we shall refer to the above system as the stochastic volatility 
  (abbreviated as SV) model. In Sections \ref{sec:modelling} and \ref{sec:expansion}, 
  we will not specify the concrete form of $f$, but assume that $f$ is bounded and 
  smooth enough, e.g., $f\in C_0^2({\mathbb R})$. Furthermore, $f$ has to satisfy a 
  sufficient growth condition 
  in order to avoid bad behavior such as the non-existence of moments of $\{S_t:t\geq 0 \}$. 
  For numerical results in Section \ref{sec:numerical}, we choose $f$ to take a special 
  form as used in \cite{fou00}, \cite{fou11} and \cite{Cao21}.
  
  We apply the well-known Girsanov theorem to change the physical measure  $\mathbb P$
  to a risk-neutral martingale measure $\mathbb Q$ by letting
  \[
  dW_t^{s*}  =\frac{\mu-r}{f\left(Y_t\right)} dt+dW^s_ty \quad \mbox{and} \quad
  dW^{y*}_t  =\xi \left(Y_t\right)dt + dW^y_t,
  \]
  where $\xi \left(Y_t\right)$ represents the premium of volatility risk. Then the model 
  equations under the measure $\mathbb{Q}$ can be written as
  \begin{eqnarray}\label{eqn:modeling_frameQ}
  dS_t &=& rS_t dt+f(Y_t)S_t dW^{s*}_t,\nonumber\\
  dY_t &=& \left[\alpha\left(m-Y_t\right)-\beta\left(\rho \frac{\mu-r}{f(Y_t)} +
  \xi(Y_t)\sqrt{1-\rho^2}\right)\right]dt\\ 
       && + \beta \left(\rho dW^{s*}_t + \sqrt{1-\rho^2} dW^{y*}_t\right).\nonumber 
  \end{eqnarray}
  Note that $\left\{W_t^{s*}:t\geq 0 \right\}$ and $\left\{W_t^{y*}:t\geq 0 \right\}$ are 
  independent standard Brownian motions under $\mathbb Q$. As an Ornstein-Uhlenbeck (OU)
  process, $\{Y_t: t\ge 0 \}$ in Eq. \eqref{eqn:modeling_frame} has an invariant 
  distribution, which is normal with mean $m$ and variance $\beta^2/2\alpha$. 
  Thus, we can expect that if mean reversion is very fast, i.e., $\alpha$ goes to 
  infinity, the process $\{S_t: t \ge 0\}$ should be close to a geometric Brownian motion. 
  This means that if mean reversion is extremely fast, then the model of Black and Scholes
  would become a good approximation. In reality, however, it may not be the case. For fast 
  but not extremely fast mean-reversion, the Black-Scholes model needs to be 
  corrected to account for the random characteristics of the volatility of a risky asset.
  For this purpose, we introduce another small parameter $\epsilon$ defined by 
  $\epsilon=1/\alpha$ as done by \cite{fou00}. For notational convenience, we put $\nu=\beta/\sqrt{2\alpha}$. With the help of these notations, the model 
  equations under $\mathbb Q$ is re-written as 
  \begin{eqnarray*}
  dS_t &=& r S_t dt+f\left(Y_t\right)S_t dW^{s*}_t,\\
  dY_t &=& \left[\frac{1}{\epsilon}\left(m-Y_t\right)-\frac{\sqrt{2} \nu}{\sqrt{\epsilon}}
  \Lambda\left(Y_t\right)\right]dt+\frac{\sqrt{2} \nu}{\sqrt{\epsilon}} dW^{y*}_t, 
  \end{eqnarray*}
  where $\Lambda(\cdot)$, defined by
  \[
  \Lambda(y) := \rho \frac{\mu-r}{f(y)} + \xi(y)\sqrt{1-\rho^2},
  \]
  is the combined market price of risk.
  
  \subsection{Path-dependent options}
  Let $V(T)$ denote the payoff of a put option on the risky asset at its expiration 
  $T$. Then its risk-neutral price at time $t \in [0, T]$ under our SV model 
  is given by
  \begin{eqnarray*}
  &&P\left( t,s,y\right)={\mathbb E}^{\mathbb Q}\left(e^{-r\left(T-t\right)}V(T)|\ 
  S_t= s, Y_t=y\right).
  \end{eqnarray*}
  Note that $V(T)$ varies depending on the type of options. In this paper, we 
  consider two types of path-dependent options: down-and-out put options and 
  floating strike lookback put options. For notational convenience, we put
  $U_t := \min_{0\le u \le t} S_u$ and $Z_t := \max_{0\le u \le t} S_u$.
  The payoff of a down-and-out put option is given by
  \[
  DOP(T) := \max\{K-S_T, 0\} \times {\mathbbm 1}_{U_T > B},
  \]
  where $K$ is the strike price, $B$ is the barrier level satisfying $0<B<K$ and
  ${\mathbbm 1}_{U_T > B}$ is the indicator function. For a floating strike 
  lookback put option, its payoff has the following form:
  \[
  LP_{float}(T) := Z_T -S_T.
  \]
  
  Applying It$\hat{\rm o}$'s lemma, we can obtain a partial differential equation
  (PDE) for $P(t,s,y)$ as follows:
  \begin{eqnarray}\label{eqn:pdeP}
  0&=& \frac{\partial P}{\partial t} + \frac{1}{2}s^2 f^2(y)
  \frac{\partial^2 P}{\partial s^2} + r\left(s\frac{\partial P}{\partial s}
  -P \right) +\frac{\sqrt{2}\rho \nu s}{\sqrt{\epsilon}} f(y) \frac{\partial^2 P}
  {\partial s\partial y} \nonumber\\
  && + \frac{\nu^2}{\epsilon} \frac{\partial^2 P}
  {\partial y^2}+ \left(\frac{1}{\epsilon}(m-y)-\frac{\sqrt{2}\nu}{\sqrt{\epsilon}} 
  \Lambda(y)\right) \frac{\partial P}{\partial y}. 
  \end{eqnarray} 
  The boundary conditions for Eq.~\eqref{eqn:pdeP} vary depending on the type of 
  options. For example, the boundary conditions for Eq.~\eqref{eqn:pdeP} when $V(T)=
  DOP(T)$ are
  \[
  \left\{
  \begin{array}{ll}
  P(T, s, y) = \max\{K-s, 0\}, & s >B, \\[0.5em]
  P(t, B, y) = 0, & 0\le t \le T.
  \end{array}
  \right.
  \]
  When $V(T)= LP_{float}(T)$, the boundary conditions become the following 
  \[
  \left\{
  \begin{array}{ll}
  {\displaystyle \frac{\partial P}{\partial z}}(t, z, y, z) = 0, & 0\le t \le T, 
  z >0, \\[0.8em]
  P(T, s, y, z) = z-s, & 0\le s \le z.
  \end{array}
  \right.
  \]
  Note that in this case, $P$ is a function of four variables $t$, $s$, $y$ and $z$
  (here, $Z_t=z$). 
  
  \section{Asymptotic Expansions} \label{sec:expansion}
  
  In this section, we apply an asymptotic expansion approach to establish partial 
  differential equations, which will be used to derive an approximate solution to
  Eq. \eqref{eqn:pdeP} and thus find an approximated value of a put option. 
   
  \subsection{Asymptotic expansions}
  We begin with re-organizing Eq.~\eqref{eqn:pdeP} in terms of the orders of $\epsilon$ 
  as follows:  
  \begin{eqnarray} \label{eqn:pdeP2}
  \frac{1}{\epsilon} {\mathcal L_0}{P} + \frac{1}{\sqrt{\epsilon}}{\mathcal L_1}
  {P} + {\mathcal L_2}{P} = 0,
  \end{eqnarray}
  where the operators ${\mathcal L_0}$, ${\mathcal L_1}$ and ${\mathcal L_2}$ are defined by
  \begin{eqnarray*}
  \mathcal L_0&:=&\left(m-y\right)\frac{\partial}{\partial y}+\nu^2 \frac{\partial^2}
  {\partial y^2},\\
  \mathcal L_1&:=&\sqrt{2}\rho \nu s f\left(y\right) \frac{\partial^2}{\partial s 
  \partial y}-\sqrt{2}\nu \Lambda\left(y\right)\frac{\partial}{\partial y}, \mbox{ and }\\
  \mathcal L_2&:=&\frac{\partial}{\partial t}+\frac{1}{2} s^2 f^2\left(y\right) 
  \frac{\partial^2}{\partial s^2}+r\left(s\frac{\partial}{\partial s}-\cdot \right).
  \end{eqnarray*}
  In order to obtain an efficient approximate solution to $P$, as that in \cite{fou06}
  and \cite{fou11}, we apply the following asymptotic expansion of $P$ as terms with 
  varying orders of $\epsilon$:
  \begin{eqnarray} \label{eqn:P-expansion}
  P=P_{0}+\sqrt{\epsilon}P_{1}+\epsilon P_{2}+\epsilon\sqrt{\epsilon}P_{3}+
  \cdots,
  \end{eqnarray}
  where $P_{0}$, $P_{1}$, ... are functions corresponding to varying orders of $\epsilon$.
  Substituting $P$ in Eq. \eqref{eqn:P-expansion} into the Eq.\eqref{eqn:pdeP2} and 
  re-organizing terms, we obtain 
  \begin{eqnarray} \label{eqn:asym_expan1}
  0 &=& \frac{1}{\epsilon}{\mathcal L_0}P_{0} + \frac{1}{\sqrt{\epsilon}}\left(\mathcal 
  L_{1}P_{0} + \mathcal L_{0}P_{1}\right) + \left(\mathcal L_{0}P_{2}+\mathcal L_{1}P_{1} 
  + \mathcal L_{2}P_{0}\right) \nonumber\\
  && +\sqrt{\epsilon}\left(\mathcal L_{0}P_{3} + \mathcal L_{1}P_{2} + 
  \mathcal L_{2}P_{1}\right) + \cdots.
  \end{eqnarray}
  Our aim is to find $P_0$ and $P_1$. 
  
  Firstly, from the $O(1/\epsilon)$-order term in Eq.\eqref{eqn:asym_expan1}, 
  we get $\mathcal L_0P_{0}=0$. If we assume that $P_{0}$ does not grow as fast as 
  $e^{y^2/2}$, we can show that $P_{0}$ is independent of $y$. Secondly, from the 
  $O(1/\sqrt \epsilon)$-order term in Eq. \eqref{eqn:asym_expan1}, we can get
  \begin{eqnarray*}
  \mathcal L_{1}P_{0}+\mathcal L_{0}P_{1} = 0.
  \end{eqnarray*} 
  Since $P_{0}$ is independent of $y$, then $\mathcal L_{1}P_{0}=0$. It follows that
  $\mathcal L_{0}P_{1}=0$. Again, if we assume that $P_{1}$ does not grow as fast as 
  $e^{y^2/2}$, then we can deduce that $P_{1}$ is also independent of $y$.
  
  Next, from the $O(1)$-order term in Eq. \ref{eqn:asym_expan1}, we get
  \[
  \mathcal L_{0}P_{2}+\mathcal L_{1}P_{1}+\mathcal L_{2}P_{0}=0.
  \]
  Since $P_1$ is independent of $y$, we have $\mathcal L_{1}P_{1}=0$ which implies that
  \begin{eqnarray} \label{eqn:asym_p_zero}
  \mathcal L_{0}P_2 + \mathcal L_{2}P_0 = 0.
  \end{eqnarray}
  Seeing Eq. \eqref{eqn:asym_p_zero} as a Poisson equation for $P_2$ in $y$, in order 
  for it to have a solution, it is required to satisfy the centring condition
  \begin{eqnarray} \label{eqn:l2p0}
  \langle \mathcal L_2 P_0 \rangle = \langle \mathcal L_{2}\rangle P_0=0,
  \end{eqnarray}
  which is equivalent to
  \begin{eqnarray} \label{eqn:pde_p0}
  \frac{\partial P_0}{\partial t}+rs\frac{\partial P_0}{\partial s}+\frac{1}{2}s^2 
  \langle f^2\rangle \frac{\partial^2 P_0}{\partial s^2}-rP_0=0.
  \end{eqnarray} 
  This is an equation for us to determine $P_0$ term.
  Here, $\langle \cdot \rangle$ denotes the expectation with respect to the invariant 
  distribution of the process $\{Y_t: t \ge 0\}$, i.e.,
  \[
  \langle h \rangle = \int_{-\infty}^{+\infty} h(y) \Phi(y)dy,
  \]
  where 
  \[
  \Phi(y) = \frac{1}{\sqrt{2\pi\nu^2}} e^{-\frac{(y-m)^2}{2\nu^2}}.
  \]
  Note that small $\epsilon$ value corresponds to fast-mean reverting. In this case,
  $Y_t$ approaches to a constant and $\langle f^2\rangle$ can be regarded as constant
  variance and then Eq. \eqref{eqn:pde_p0} is the Black-Scholes PDE. Thus, for small
  $\epsilon$, $P_0$ represents the put option price under the Black-Scholes model. 
  
  Following Eq. \eqref{eqn:l2p0}, we have
  \begin{eqnarray*}
  \mathcal L_2P_0 =\mathcal L_2P_0 - \langle \mathcal L_2 \rangle P_0 = 
  \frac{1}{2}\left(f^2- \langle f^2\rangle\right) s^2 \frac{\partial^2 P_0}{\partial s^2},
  \end{eqnarray*}
  which together with Eq. \eqref{eqn:asym_p_zero} implies 
  \begin{eqnarray} \label{eqn:p-two}
  \mathcal L_0P_2 = -\frac{1}{2} \left(f^2-\langle f^2\rangle\right) s^2 
  \frac{\partial^2 P_0}{\partial s^2}.
  \end{eqnarray}
  The solution to Eq. \eqref{eqn:p-two} can be expressed as
  \begin{eqnarray} \label{eqn:p2-solution}
  P_2 = -\frac{1}{2}\left(\phi+ c\right) s^2 \frac{\partial^2 P_0}{\partial s^2},
  \end{eqnarray}
  where $\phi$ is a function of $y$ which only satisfies the equation 
  \[
  {\mathcal L}_0\phi = f^2 -\langle f^2\rangle
  \] 
  and $c$ is a function of other variables except $y$.
  
  To derive an equation for $P_1$, we consider the $O(\sqrt{\epsilon})$-term
  in Eq. \eqref{eqn:asym_expan1} and obtain
  \[
  {\mathcal L}_0 P_3 + {\mathcal L}_1 P_2 + {\mathcal L}_2 P_1 = 0.
  \]
  This equation can be regarded as a Poisson equation for $P_3$ in $y$, and in order for
  it to have a solution, the following centring condition must be satisfied:
  \begin{eqnarray} \label{eqn:p1-equation1}
  \langle\mathcal L_1P_2+\mathcal L_2P_1\rangle =0.
  \end{eqnarray}
  After we substitute $P_2$ in Eq. \eqref{eqn:p2-solution} into Eq. \eqref{eqn:p1-equation1}
  and make simplification, we obtain 
  \begin{eqnarray} \label{eqn:p1-equation2}
  \frac{\partial P_1}{\partial t}+\frac{1}{2}\langle f^2 \rangle s^2 \frac{\partial^2 P_1}
  {\partial s^2}+rs\frac{\partial P_1}{\partial s}-rP_1 =
  c_1 s^3 \frac{\partial^3 P_0}{\partial s^3}
  + c_2 s^2 \frac{\partial^2 P_0}{\partial s^2},
  \end{eqnarray}
  where
  \begin{eqnarray} \label{eqn:c1andc2}
  c_1 := \frac{\sqrt{2}}{2}\langle f\phi'\rangle \rho\nu \quad \mbox{and} \quad 
  c_2 := \frac{\sqrt{2}}{2}\left(2 \rho \langle f \phi'\rangle - \langle \Lambda \phi'
  \rangle\right) \nu.
  \end{eqnarray}
  This is an equation for us to determine the first correction term $P_1$.
  
  We summarize the key points in the previous formal analysis as the following theorem. 
  
  \begin{theorem} \label{thm:general-value}
  Under the SV model governed by Eq. \eqref{eqn:modeling_frame}, the risk-neutral value 
  $P$ of a path-dependent put option can be approximated by the following formula
  \begin{eqnarray}
  P \approx P_0 + \sqrt{\epsilon} P_1,
  \end{eqnarray}
  for small $\epsilon$, where $P_0$ and $P_1$ are determined by Eq. \eqref{eqn:pde_p0} 
  and Eq. \eqref{eqn:p1-equation2} with corresponding boundary conditions, respectively.
  $P_0$ is the put option price under the Black-Scholes model with constant effective
  volatility $\sqrt{\langle f^2 \rangle}$ and $P_1$ is the first-order correction term.
  \end{theorem}
  
  Finally, as mentioned in Section \ref{sec:modelling}, boundary conditions for 
  Eq. \eqref{eqn:l2p0} and Eq. \eqref{eqn:p1-equation2} depend on the types 
  of options we consider. We describe the corresponding boundary conditions and 
  solve these equations in the next two sections. 
  
  \section{Solving $P_0$ and $P_1$ for Down-and-out Put Options} \label{sec:barrier}
  
  In this section, we use Mellin transform to derive analytical expressions of the $P_0$ 
  and $P_1$ terms for down-and-out put options.
  
  \subsection{$P_0$ term for down-and-out put options}
  In order to use Mellin transform to calculate the $P_0$ term for down-and-out put options, 
  noting that $P_0$ is independent of $y$ under our assumption, we first follow the method 
  in \cite{buc11} and use the boundary condition, 
  \[
  P(T, s, y) =\max\{K-s, 0\}, \quad \mbox{for} \quad s >B, 
  \]
  to set up the boundary condition of $P_0$ for $s\ge 0$ as follows:
  \begin{eqnarray} \label{eqn:boundary_exten}
  P_0 \left(T, s \right):=\left(K- s\right) {\mathbbm 1}_{B<s<K}-\left(\frac{B}{s} 
  \right)^{k_1-1} \left( K-\frac{B^2}{s}\right) {\mathbbm 1}_{\frac{B^2}{K}<s<B},
  \end{eqnarray}
  where $k_1 = 2r/\langle f^2 \rangle$.  Now, we apply Mellin transform to 
  Eq. \eqref{eqn:pde_p0} to convert this PDE into the following ODE: 
  \begin{eqnarray} \label{eqn:ODE_P0}
  \frac{d \hat P_0}{dt}+ \left(\frac{1}{2}\langle f^2 \rangle(w^2+w)-rw-r\right)\hat P_0=0.
  \end{eqnarray}
  The solution to Eq. \eqref{eqn:ODE_P0} is given by
  \begin{eqnarray} \label{eqn:ODE_P0_solu}
  \hat P_0\left(t,w\right) = \hat{\theta}(w) e^{\frac{1}{2} \langle f^2 
  \rangle\left(w^2+\left(1-k_1\right)w-k_1\right)\left(T-t\right)},
  \end{eqnarray}
  where $\hat{\theta}$ is a function of $w$, determined by the boundary condition
  \eqref{eqn:boundary_exten}. 
  
  Next, we take inverse Mellin transform of Eq. \eqref{eqn:ODE_P0_solu} and obtain
  \[
  P_0(t,s) = P_0(T,s) * {\mathcal M}^{-1} e^{\lambda\left(w+\eta\right)^2+\delta},
  \]
  where
  \[
  \lambda=\frac{1}{2} \langle f^2 \rangle\left(T-t\right),\ \eta=\frac{1-k_1}{2},\  
  \delta =-\lambda \eta^2-r\left(T-t\right)
  \]
  and the operation $*$ means the convolution. Applying Table 1 in Appendix \ref{app:mellin} 
  and the boundary condition given in Eq. \eqref{eqn:boundary_exten}, we have

  \begin{eqnarray} \label{eqn:down-p0-mellin}
  P_0\left(t, s\right)
  &=&P_0\left(T,s\right)*\left(\frac{e^{\delta} s^{\eta}}{2\sqrt{\lambda\pi}} 
  e^{-\frac{1}{4\lambda} \left(\ln s\right)^2}\right) \nonumber\\
  &=&\int_{B}^{K} \left(K-u\right)e^{\delta} \left(\frac{s}{u}\right)^{\eta} 
  \left(\frac{1}{2\sqrt{\lambda\pi}} e^{-\frac{1}{4\lambda}
  \left(\ln \left(\frac{s}{u}\right)\right)^2} \right)\frac{du}{u}-\\[0.5em]
  &&\int_{\frac{B^2}{K}}^{B} \left(\frac{B}{u} \right)^{k_1-1} \left(K-\frac{B^2}{u}
  \right) e^{\delta} \left(\frac{s}{u}\right)^{\eta} \left(\frac{1}{2\sqrt{\lambda\pi}} 
  e^{-\frac{1}{4\lambda}\left(\ln \left(\frac{s}{u}\right)
  \right)^2} \right)\frac{du}{u}. \nonumber
  \end{eqnarray}
  After some careful calculation, for down-and-out put options, we derive a closed-form 
  expression of the $P_0$ term as follows:
  \begin{eqnarray}\label{eqn:down-put-p0}
  P_0(t, s)&=&K e^{-r\left(T-t\right)} \left(\Phi\left(-\Delta_{-}\left(\frac{s}{K}
  \right)\right)-\Phi\left(-\Delta_{-}
  \left(\frac{s}{B}\right)\right)\right) -\nonumber\\
  &&s\left(\Phi\left(-\Delta_{+}\left(\frac{s}{K}\right)\right)-\Phi
  \left(-\Delta_{+}\left(\frac{s}{B}\right)\right) \right) -\nonumber\\
  &&K e^{-r\left(T-t\right)} \left(\frac{B}{s}\right)^{k_1-1} \left[\Phi
  \left(\Delta_{-}\left(\frac{B}{s}\right)\right)-\Phi\left(\Delta_{-}
  \left(\frac{B^2}{sK}\right)\right)\right] +\nonumber\\
  &&B\left(\frac{B}{s}\right)^{k_1} \left[\Phi\left(\Delta_{+}\left(\frac{B}{s}
  \right)\right)-\Phi\left(\Delta_{+}\left(\frac{B^2}{sK}\right)\right) \right],
  \end{eqnarray}
  where $\Phi(\cdot)$ is the CDF of the standard normal distribution and
  \[
  \Delta_{\pm}(x)=\frac{1}{\sqrt{\langle f^2 \rangle (T-t)}}\left[\ln (x)+\left(r\pm 
  \frac{1}{2} \langle f^2 \rangle\right)(T-t)\right]. 
  \] 
  Note that $P_0$ given in Eq. \eqref{eqn:down-put-p0} is precisely the same as the 
  price of a down-and-out put option given in the literature, e.g., \cite{Hull2015} 
  (Chapter 26, p.606) or \cite{Haug06} (Chapter 4), if we let $\sigma^2 = \langle 
  f^2 \rangle$. For details of the derivation of formula \eqref{eqn:down-put-p0}, 
  we refer the reader to Appendix \ref{app:deviation}.
  
  \subsection{$P_1$ term for down-and-out put options}
  For down-and-out put options, the boundary conditions for $P_1$ are given 
  as follows:
  \[
  \left\{
  \begin{array}{lcl}
  P_1(T,s)&=&0, \quad for\ \  s \ge B,\\[0.8em]
  P_1(t, B) & = &0, \quad for\ \ 0< t <T.
  \end{array}
  \right.
  \]
  We again follow the method in Buchen (2001) and extend the boundary conditions 
  $P_1(T, s)=0$, for $s \ge B$ as $P_1\left(T, s\right)=0$ for all $s \ge 0$.
  
  Next, we apply Mellin transform to Eq. \eqref{eqn:p1-equation2} to get
  \begin{eqnarray*} \label{eqn:Q1-lookout-Mellin}
  \frac{d\hat{P_1}}{dt}+\left(\frac{1}{2}\langle f^2 \rangle\left(w^2+w\right)-rw-
  r\right)\hat{P_1}
  =\left(-c_1 w\left(w+1\right)\left(w+2\right)+c_2w\left(w+1\right)\right)\hat{P}_0.
  \end{eqnarray*}
  Solving this equation, we obtain
  \[
  \hat{P}_1\left(t,w\right)=\left[c_1\left(T-t\right)w^3-\left(c_2-3c_1\right)
  \left(T-t\right)w^2-\left(c_2-2c_1\right)\left(T-t\right)w\right] \hat{P}_0
  \left(t,w\right).
  \]
  
  Finally, applying inverse Mellin transform, we obtain an explicit closed-form 
  expression of $P_1$ as follows
  \begin{eqnarray} \label{eqn:down-put-p1}
  P_1\left(t,s\right)&=& {\mathcal M}^{-1}\left(\hat{P}_1\left(t, w\right)\right) 
  \nonumber\\
  &=&c_1\left(T-t\right)\left(-s\frac{d}{ds}P_0\left(t,s\right)-3s^2\frac{d^2}{ds^2} 
  P_0\left(t,s\right)-s^3\frac{d^3}{ds^3}P_0\left(t,s\right)\right) \nonumber\\
  &&-\left(c_2-3c_1\right)\left(T-t\right)\left(s\frac{d}{ds}P_0\left(t,s\right)+s^2
  \frac{d^2}{ds^2} P_0\left(t,s\right)\right) \\
  &&-\left(c_2-2c_1\right)\left(T-t\right)\left(-s\frac{d}{ds} P_0\left(t,s\right)\right),
  \nonumber
  \end{eqnarray}
  where $P_0$ is given in the previous section, $c_1$ and $c_2$ are given in 
  Eq. \eqref{eqn:c1andc2}. 
  
  We summarize the above analysis and calculation on down-and-out put options in the 
  following theorem.
  
  \begin{theorem}
  Under the SV model governed by Eq. \eqref{eqn:modeling_frame}, the risk-neutral 
  value $P$ of a down-and-out put option can be approximated by the following formula
  \begin{eqnarray}
  P \approx P_0 + \sqrt{\epsilon} P_1,
  \end{eqnarray}
  where $P_0$ and $P_1$ are given by Eq. \eqref{eqn:down-put-p0} and Eq.
  \eqref{eqn:down-put-p1}, respectively.
  \end{theorem}
  
  \section{Solving $P_0$ and $P_1$ for Lookback Put Options} \label{sec:lookback}
  
  In this section, we use Mellin transform to derive analytical expressions of the 
  $P_0$ and $P_1$ terms for floating strike lookback put options.

  \subsection{$P_0$ term for lookback put options}
  For lookback floating strike put options, the boundary conditions of $P_0$ are
  \[
  \left\{
  \begin{array}{lcl}
  {\displaystyle \frac{\partial P_0}{\partial z}} (t,z,z)&=&0,\\[0.8em]
  {\displaystyle \frac{\partial P_0}{\partial z}}\left(T, s, z\right) & = &1,
  \quad for\ \ 0< s <z.
  \end{array}
  \right.
  \]
  Similar to the case of down-and-out put options, we extend the second boundary condition
  to $0<s < \infty$ as follows:
  \begin{eqnarray*}
  \frac{\partial P_0}{\partial z}\left(T, s, z\right) := {\mathbbm 1}_{s < z} -
  \left(\frac{z}{s}\right)^{k_1-1}\cdot {\mathbbm 1}_{z<s},
  \quad \mbox{for} \quad 0<s<\infty.
  \end{eqnarray*} 
  Then, by integrating each side of the last equation, we can obtain
  \begin{eqnarray} \label{eqn:extended_P_0}
  P_0\left(T, s, z\right) =\int_s^z -\left(\frac{\xi}{s}\right)^{k_1-1} 
  d\xi = -\frac{1}{k_1}\left(\frac{z}{s}\right)^{k_1} s+\frac{1}{k_1}s
  \end{eqnarray}
  for $s>z$. For convenience, we let $u=s/z$ and $Q_0 = P_0/z$. With these
  notations, Eq. \eqref{eqn:pde_p0} becomes
  \begin{eqnarray} \label{eqn:pde_q0}
  \frac{\partial Q_0}{\partial t}+\frac{1}{2}u^2 \langle f^2\rangle \frac{\partial^2 Q_0}
  {\partial u^2} + ru\frac{\partial Q_0}{\partial u} - rQ_0=0,
  \end{eqnarray} 
  with boundary conditions
  \begin{eqnarray} \label{eqn:pde_Q_0}
  Q_0\left(T, u\right) = -\frac{1}{k_1} u^{1-k_1} +\frac{1}{k_1}u, \quad \mbox{for}
  \quad u>1,
  \end{eqnarray}
  and $Q_0(T,u) = 1$, for $0<u<1$.
 
  Note that except the boundary conditions, Eq. \eqref{eqn:pde_q0} is identical to
  Eq. \eqref{eqn:pde_p0}. Applying Mellin transform in the same way as that for the 
  case of down-and-out put options, we can derive the solution to Eq. \eqref{eqn:pde_q0} 
  as follows:
  \[
  Q_0(t,u)=\hat{\theta}(w)*{\mathcal M}^{-1} e^{\lambda\left(w+\eta\right)^2+\delta}.
  \]
  Again, applying Table 1 and $P_0$ given in Eq. \eqref{eqn:boundary_exten}, we have
  \begin{eqnarray} \label{eqn:floating-q0-mellin}
  Q_0(t, u) &=& Q_0(T,u) *e^{\delta} z^{\eta}\left(\frac{1}{2\sqrt{\pi}} 
  \lambda^{-\frac{1}{2}} e^{-\frac{1}{4\lambda} \left(\ln z\right)^2}\right)
  \nonumber\\[0.5em]
  &=&\int_{0}^{1} \left(1-\xi\right)e^{\delta} \left(\frac{u}{\xi}\right)^{\eta} 
  \left(\frac{1}{2\sqrt{\pi}} \lambda^{-\frac{1}{2}} e^{-\frac{1}{4\lambda}
  	\left(\ln \left(\frac{u}{\xi}\right)\right)^2} \right)\frac{d\xi}{\xi} + \\[0.5em]
  &&\int_{1}^{\infty} \left(\frac{-1}{k_1}\xi^{1-k_1} +\frac{\xi}{k_1} \right)e^{\delta} 
  \left(\frac{u}{\xi}\right)^{\eta} \left(\frac{1}{2\sqrt{\pi}} \lambda^{-\frac{1}{2}} 
  e^{-\frac{1}{4\lambda}\left(\ln \left(\frac{u}{\xi}\right)\right)^2} \right)
  \frac{d\xi}{\xi}. \nonumber
  \end{eqnarray}
  After calculating integrals, for floating strike lookback put options, we derive a 
  closed-form expression of the $P_0$ term as follows:
  \begin{eqnarray}\label{eqn:floating-put-p0}
  P_0(t,s, z)&=& z e^{-r\left(T-t\right)}\Phi\left(-\Delta_{-}\left(\frac{s}{z}\right)
  \right)-s \Phi\left(-\Delta_{+}\left(\frac{s}{z}\right)\right) \\[0.5em]
  && -\frac{z}{k_1} \left(\frac{s}{z}\right)^{1-k_1} e^{-r\left(T-t\right)} \Phi\left(-\Delta_{-}\left(\frac{z}{s}\right)\right)
  +\frac{s}{k_1} \Phi\left(\Delta_{+}\left(\frac{s}{z}\right)
  \right), \nonumber
  \end{eqnarray}
  where $\Phi(\cdot)$ is the CDF of the standard normal distribution. Note that $P_0$ 
  given in Eq. \eqref{eqn:floating-put-p0} is precisely the same as the price of 
  a floating strike put option given in the literature, e.g., \cite{Hull2015} 
  (Chapter 26, p.608) or \cite{Haug06} (Chapter 4), if we let $\sigma^2 := \langle 
  f^2 \rangle$. Details of the derivation of this formula can be found in Appendix \ref{app:deviation}.
  
  \subsection{$P_1$ term for lookback put options}
  For lookback floating strike put options, the boundary conditions for $P_1$ are given 
  as follows:
  \[
  \left\{
  \begin{array}{lcl}
  P_1(T,s,z)&=&0, \quad for\ \ 0< s <z,\\[0.8em]
  {\displaystyle \frac{\partial P_1}{\partial z}}\left(t, z, z\right) & = &0,
  \quad for\ \ 0< t <T \quad and\ z>0.
  \end{array}
  \right.
  \]
  
  Just like that for the $P_0$-term for floating strike lookback put options, we let 
  $u=s/z$ and $Q_1 = P_1/z$.
  With these notation changes, Eq. \eqref{eqn:p1-equation2} is converted to the following
  \begin{eqnarray} \label{eqn:Q1-lookout}
  \frac{\partial Q_1}{\partial t}+\frac{1}{2}\langle f^2 \rangle u^2 \frac{\partial^2 Q_1}
  {\partial u^2}+ru\frac{\partial Q_1}{\partial u}-rQ_1=c_1 u^3\frac{\partial^3 Q_0}
  {\partial u^3}+c_2 u^2\frac{\partial^2 Q_0}{\partial u^2}
  \end{eqnarray}
  with $Q_1(T, u)=0$ for $0 < u < 1$.
  
  Note that Eq. \eqref{eqn:Q1-lookout} is essentially the same as Eq. \eqref{eqn:p1-equation2}, 
  except the notational difference. So, we have
  \begin{eqnarray}
  Q_1\left(t,u \right)&=& 
  c_1\left(T-t\right)\left(-u\frac{d}{du}Q_0\left(t,u\right)-3u^2\frac{d^2}{du^2} Q_0
  \left(t,u\right)-u^3\frac{d^3}{du^3}Q_0\left(t,u\right)\right) \nonumber\\
  &&-\left(c_2-3c_1\right)\left(T-t\right)\left(u\frac{d}{du}Q_0\left(t,u\right)+u^2
  \frac{d^2}{dz^2} Q_0\left(t,u\right)\right) \\
  &&-\left(c_2-2c_1\right)\left(T-t\right)\left(-u\frac{d}{du} Q_0\left(t,u\right)\right),
  \nonumber
  \end{eqnarray}
  where $Q_0$ is given previously. Consequently, we have
  \begin{eqnarray} \label{eqn:floating-put-p1}
  P_1\left(t,s,z\right)
  &=&c_1\left(T-t\right)\left(-s\frac{d}{ds}P_0\left(t,s,z\right)-3s^2\frac{d^2}{ds^2} P_0
  \left(t,s,z\right)-s^3\frac{d^3}{ds^3}P_0\left(t,s,z\right)\right) \nonumber\\
  &&-\left(c_2-3c_1\right)\left(T-t\right)\left(s\frac{d}{ds}P_0\left(t,s,z\right)
  +s^2 \frac{d^2}{ds^2} P_0\left(t,s,z\right)\right)\\
  &&-\left(c_2-2c_1\right)\left(T-t\right)\left(-s\frac{d}{ds} P_0\left(t,s,z\right)\right),
  \nonumber
  \end{eqnarray}
  where $c_1$ and $c_2$ are the same as those defined previously.
  
  We summarize the above analysis and calculation on floating strike lookback put options 
  in the following theorem.
  
  \begin{theorem}
  Under the SV model governed by Eq. \eqref{eqn:modeling_frame},
  the risk-neutral value $P$ of a floating strike lookback  put option can be approximated 
  by the following formula
  \begin{eqnarray}
  P \approx P_0 + \sqrt{\epsilon} P_1,
  \end{eqnarray}
  where $P_0$ and $P_1$ are given by Eq. \eqref{eqn:floating-put-p0} and Eq.
  \eqref{eqn:floating-put-p1}, respectively.
  \end{theorem}
  
  \section{Numerical Results and Sensitivity Analysis} \label{sec:numerical}
  
  In this section, we conduct a numerical study to investigate the sensitivity of the 
  first-order correction term $P_1$ and our approximation results $P_0 +\sqrt{\epsilon}
  P_1$ with respect to the initial value of underlying asset. This means that we set 
  $t=0$ throughout this section. We also compare the results given by our closed form 
  formulas with those generated by the Monte-Carlo simulation. 
  
  First of all, as done by \cite{fou00}, \cite{fou11} and \cite{Cao21}, we choose $f$ 
  to take the following form  
  \[
  f(y)= 0.35 \left(\tan^{-1}(y) + \frac{\pi}{2}\right)/\pi +0.05.
  \]
  Secondly, the values of other parameters used in this section are given in Table 
  \ref{table1}, whenever they are required to be fixed.
  \begin{table}[hbt!] 
  \centering
  \caption{The role and numerical value of parameters.}
  \begin{tabular}{|c|l|l|}   \hline
  		Parameter & Role & Value  \\	\hline
  		$r$ & risk-free interest rate & 0.035 \\ 
  		$B$ & barrier level & 1500 \\ 
  		$K$ & put option strike price & 2700 \\ 
  		$c_1$ & as defined in Section \ref{sec:expansion} & -0.004 \\ 
  		$c_2$ & as defined in Section \ref{sec:expansion} & -0.018 \\ \hline 
  \end{tabular}
  \label{table1}
  \end{table}
  
  \noindent
  Here, we do not choose precise values of $\beta$ and $\rho$, and particular forms 
  of $\xi(y)$ (in Section \ref{sec:modelling}) and $\phi(y)$ (in Section 
  \ref{sec:expansion}) to calculate the above values of $c_1$ 
  and $c_2$. Instead, $c_1$ and $c_2$ are calibrated from the term structure of the 
  implied volatility surface as described in the book of \cite{fou00}. 
  Specifically, the implied volatility $I^{\epsilon}$ of a European vallina call 
  option with fast mean-reverting stochastic process can be approximated by the 
  following formula
  \[
  I^{\epsilon} = a \frac{\ln (\frac{K}{s})}{T-t}+b+o(\sqrt{\epsilon})
  \]
  with
  \[
  a =-\frac{c_1}{{\langle f^2\rangle}^{3/2}} \quad \mbox{and} \quad
  b =\sqrt{\langle f^2\rangle}+\frac{c_1}{{\langle f^2\rangle}^{3/2}} \left(r+\frac{3}{2} 
  {\langle f^2\rangle} \right)-\frac{c_2}{\sqrt{\langle f^2\rangle}}.
  \]
  The parameters $a$ and $b$ are estimated as the slope and intercept of the regression 
  fit of the observed implied volatilities as a linear function of 
  logmoneyness-to-maturity-ratio $\ln (K/s)/(T-t)$. From the calibrated 
  values $a$ and $b$ on the observed implied volatility surface, the parameters $c_1$ 
  and $c_2$ are obtained as
  \[
  c_1=-a{\sigma\langle f^2\rangle}^{3/2} \quad \mbox{and} \quad
  c_2=\sqrt{\langle f^2\rangle}((\sqrt{\langle f^2\rangle}-b)-a(r+\frac{3}{2} 
  {\langle f^2\rangle})).
  \]
  Thirdly, note that when $t=0$, $s=z$. Hence, in this case, the formula for $P_0$
  given by Eq. \eqref{eqn:floating-put-p0} is simplified.
  
\begin{figure}[H]
  \centering
  \subfloat[]{\includegraphics[width=6cm]{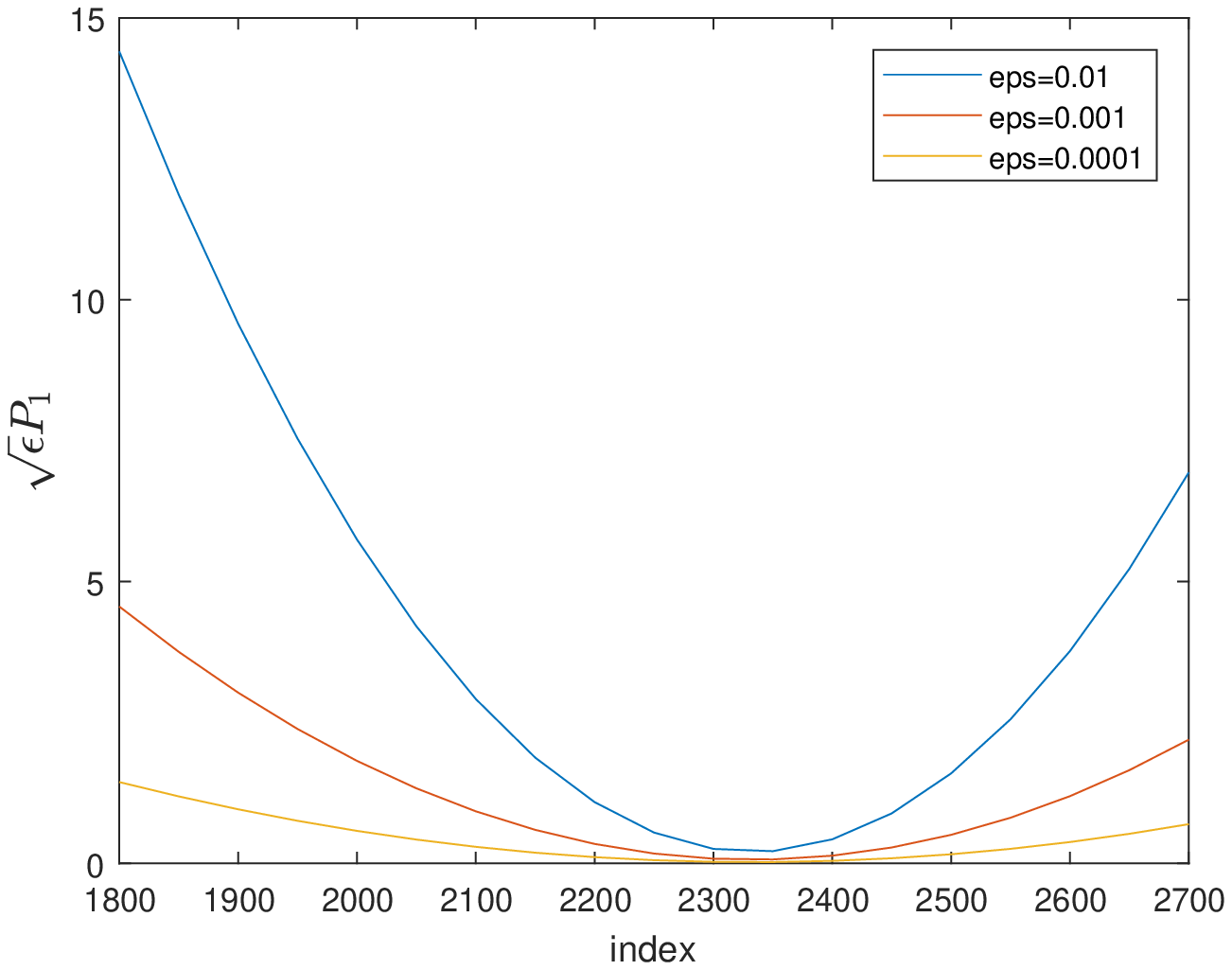}\label{Fig:DAO-p1}}
  \subfloat[]{\includegraphics[width=6cm]{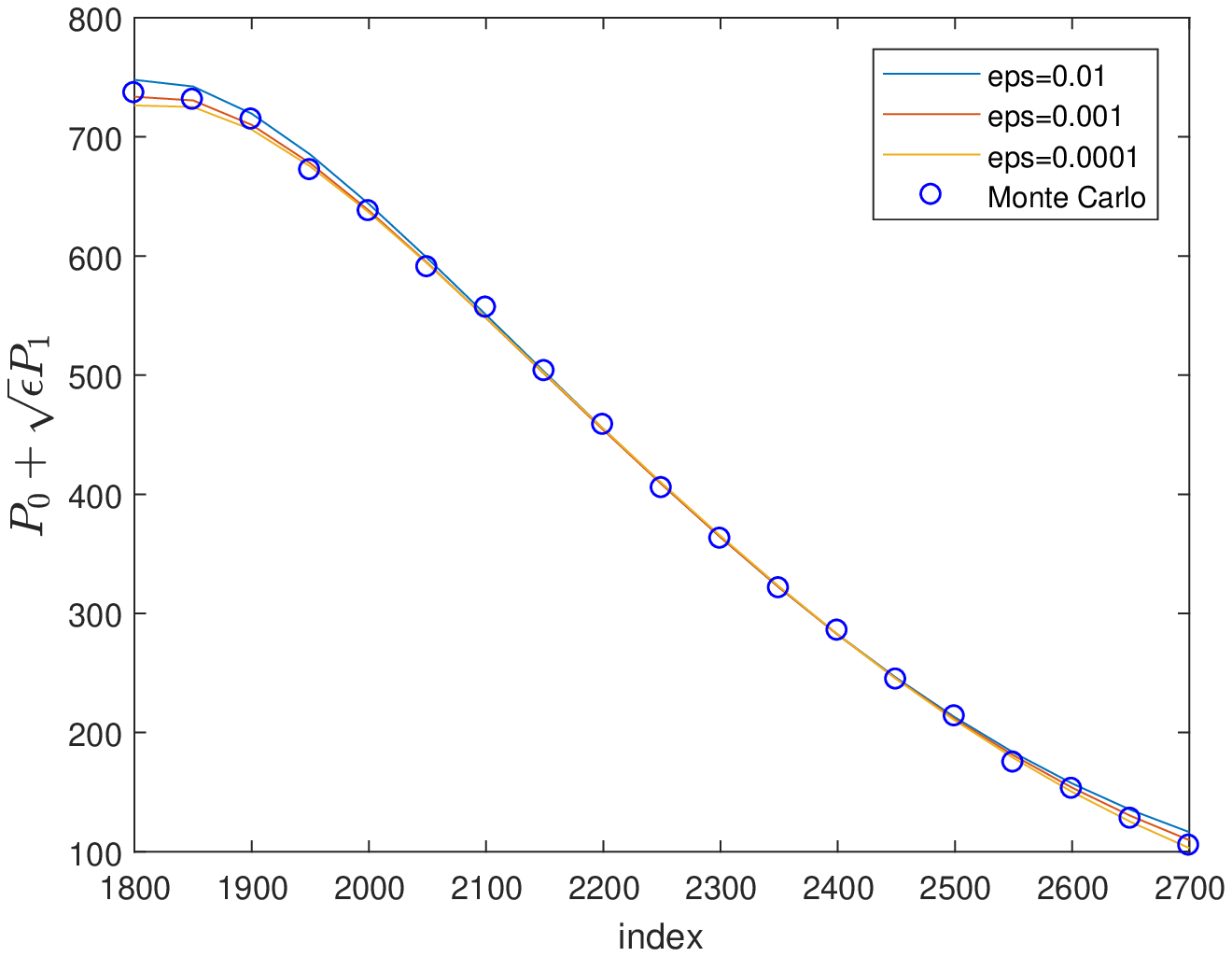}\label{Fig:DAO-approx}}
  \caption{Plots of $\sqrt{\epsilon}P_1$ and $P_0+\sqrt{\epsilon} P_1$ against 
  different values of $\epsilon$ for down-and-out put option} 
  \end{figure}

  Figure \eqref{Fig:DAO-p1} shows how the $\sqrt{\epsilon}P_1$-term for a down-and-out 
  put option changes with respect to a variation of $\epsilon$ values. As we can see, 
  for fixed $\epsilon$, when $s$ increases, $P_1$ decreases first, and then increases 
  after it hits its trough. When $\epsilon$ gets smaller (equivalently, the 
  mean-reverting speed gets larger), $\sqrt{\epsilon}P_1$ approaches to a zero.
  Figure \eqref{Fig:DAO-approx} shows how the value of $P_0 + \sqrt{\epsilon} P_1$ 
  for a down-and-out put option varies with respect to 
  the change of $\epsilon$ values. As we can see, when the value of $\epsilon$ changes 
  from 0.01 to 0.0001, the value of $P_0 + \sqrt{\epsilon} P_1$ does not vary much. In 
  fact, the values of $P_0 + \sqrt{\epsilon} P_1$ match well with the result of Monte-Carlo simulation in all cases. Furthermore, in all cases, the value of $P_0 + \sqrt{\epsilon} 
  P_1$ declines as $s$ increases.
  
\begin{figure}[H]
  \centering
  \subfloat[]{\includegraphics[width=6cm]{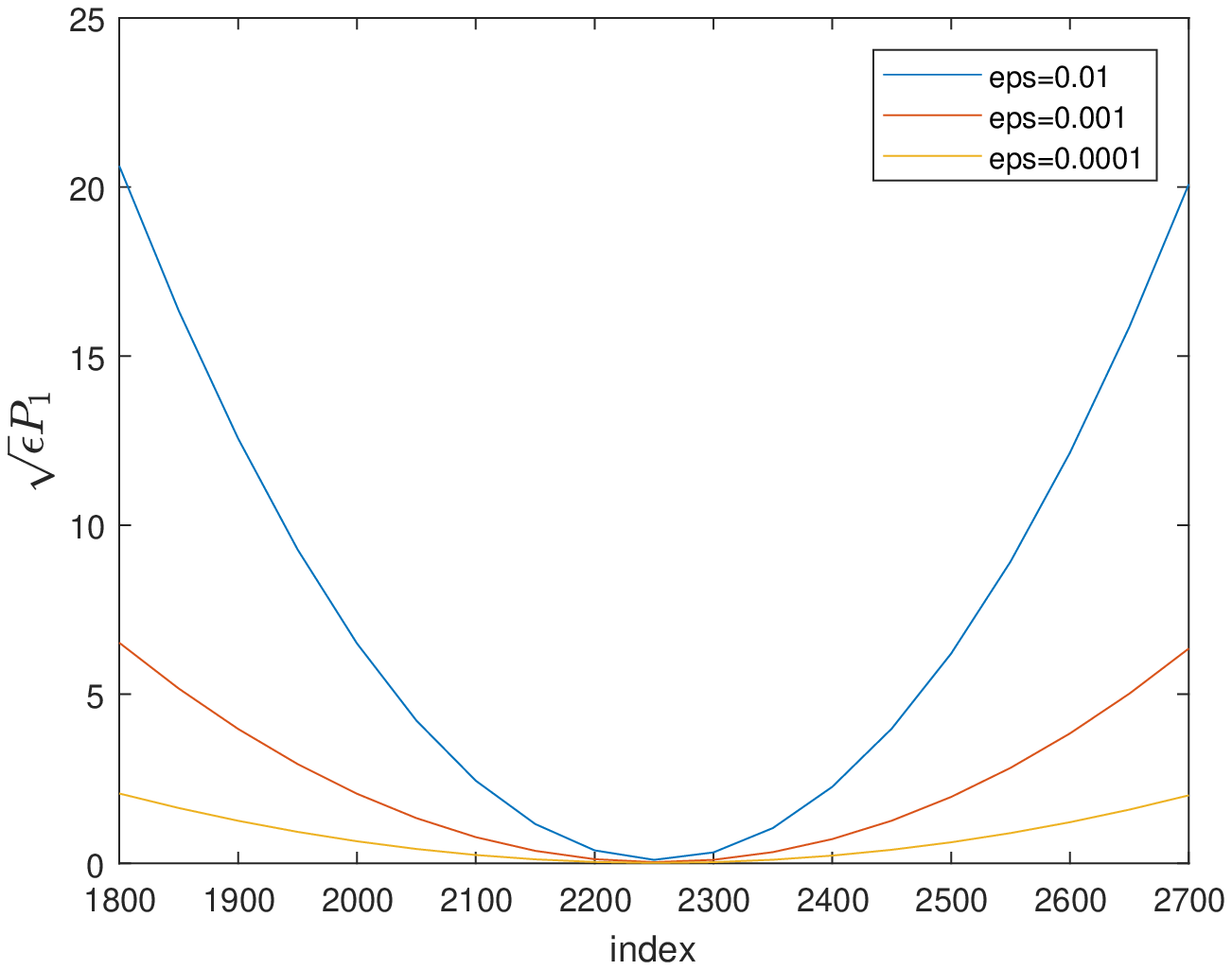} \label{Fig:lookback-p1}}
  \subfloat[]{\includegraphics[width=6cm]{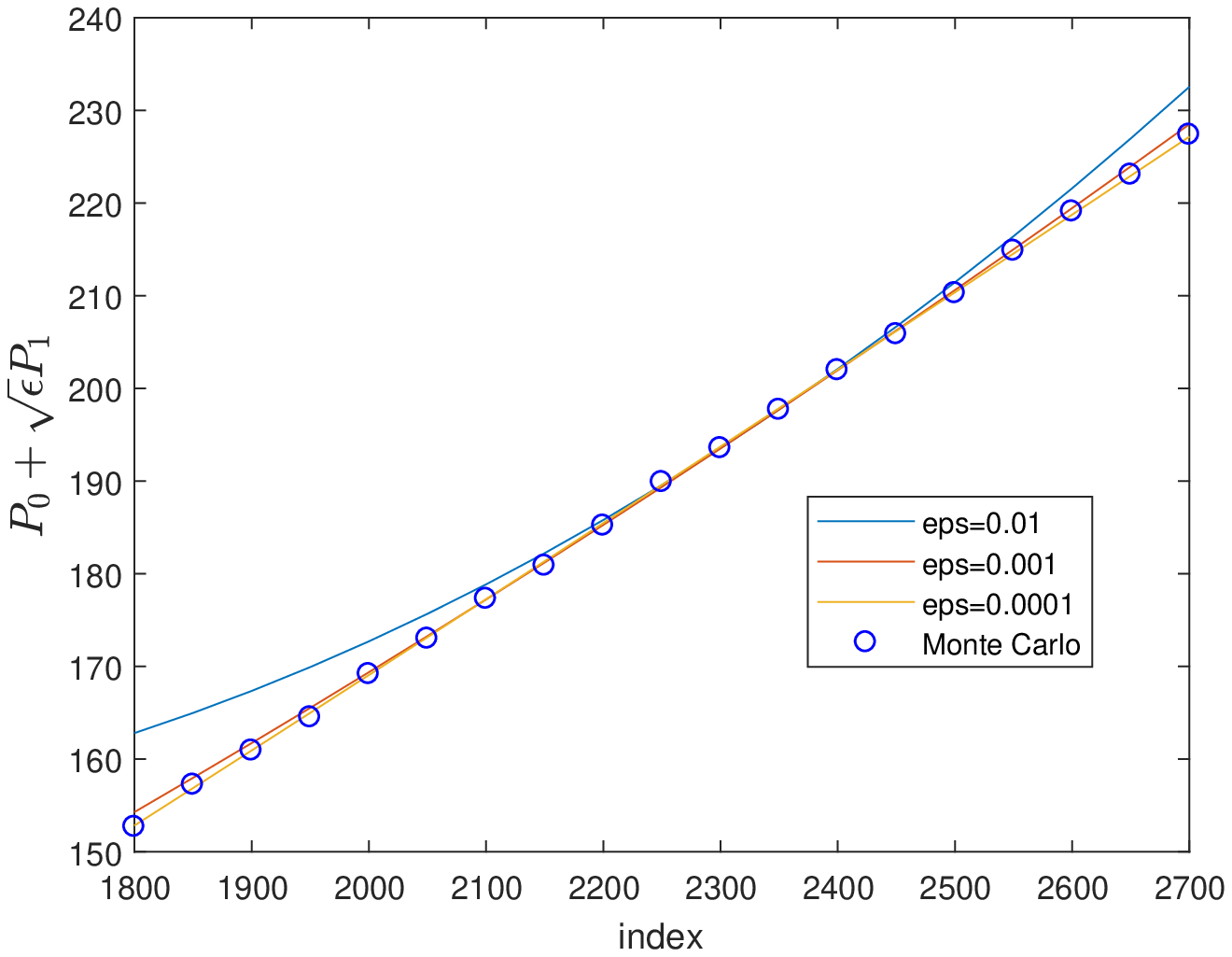}\label{Fig:lookback-approx}}
  \caption{Plots of $\sqrt{\epsilon}P_1$ and $P_0+\sqrt{\epsilon} P_1$ against different 
  values of $\epsilon$ for floating strike put option}
  \end{figure}

  Figure \eqref{Fig:lookback-p1} shows how the $\sqrt{\epsilon}P_1$-term for a floating 
  strike put changes with respect to a variation of $\epsilon$ values.  In a similar 
  pattern, for a fixed $\epsilon$-value, when $s$ increases, $P_1$ decreases first and 
  then increases after it hits its trough. Similar to the case of down-and-out put 
  options, when $\epsilon$ gets smaller (equivalently, the mean-reverting speed gets 
  larger), $\sqrt{\epsilon}P_1$ approaches to zero.
  Figure \eqref{Fig:lookback-approx} shows how the value of $P_0 + \sqrt{\epsilon} P_1$ 
  for a floating strike put varies with respect to the change of $\epsilon$ values. 
  When the value of $\epsilon$ changes from 0.01 to 0.001, the value of $P_0 + 
  \sqrt{\epsilon} P_1$ varies. But, when the value of $\epsilon$ changes from 0.001 
  to 0.0001, the value of $P_0 + \sqrt{\epsilon} P_1$ does not vary much.
  The values of $P_0 + \sqrt{\epsilon} P_1$ match well with the result of Monte-Carlo 
  simulation when $\epsilon =0.001$ or $0.0001$. Furthermore, in all cases, the value 
  of $P_0 + \sqrt{\epsilon} P_1$ increases as $s$ increases.

  \section{Conclusion Remarks}\label{sec:conclusion}	
  
  This article establishes explicit closed-form solutions for first order approximations 
  of down-and-out and floating strike put option prices under a stochastic volatility 
  model by means of Mellin transform. The zero-order terms in the solutions for the 
  prices of both types of put options coincide with those in \cite{Hull2015} or 
  \cite{Haug06} under the classical Back-Scholes model. Our numerical analysis shows
  that the results given by those explicit closed-form solutions match well with those 
  generated by Monte-Carlo simulation. This confirms the accuracy of the approximation.  
  Furthermore, we also discussed the sensitivity of the first-order error terms and the
  approximation with respect to the underlying asset price and the mean-reverting speed 
  of the OU-process which governs the volatility.
  
  \appendix
  
  \section{Mellin Transform} \label{app:mellin}
  
  The Mellin transform is an integral transform that may be 
  regarded as the multiplicative version of the two-sided Laplace transform. This integral 
  transform is often used in the theory of asymptotic expansions. For a locally Lebesgue 
  integrable function $h: \mathbb R^+ \to \mathbb R$, the Mellin transform 
  denoted by ${\mathcal M}h$ or ${\hat h}$, is given by
  \[
  {\hat h}(w) = \left({\mathcal M}h\right)(w) := \int_0^{+\infty} s^{w-1} h(s)\,ds,\quad w\in {\mathbb C},
  \]
  and if $a < {\rm Re}(w) < b$ and $c$ such that $a<c<b$ exists, the inverse of the Mellin 
  transform is expressed by
  \[
  h(s) = \left({\mathcal M}^{-1}\hat{h}\right) (s)={\frac {1}{2\pi i}}\int_{c-i\infty }^{c+i\infty}
  s^{-w}\hat{h}(w)\,dw.
  \]
  In this paper, we use the following properties of Mellin transform.
  
  \begin{table}[hbt!]
  	\centering
  	\caption{The role and numerical value of parameters.}
  	\begin{tabular}{l||l} \hline
  		\mbox{function} & \mbox{Mellin tansform}\\ \hline
  		$h$             & $\hat{h}$\\ \hline
  		$sh'$           & $-w\hat{h}$ \\ \hline
  		$s^2h''$        & $w(w+1) \hat{h}$ \\ \hline
  		$s^3 h^{(3)}$   & $-w(w+1)(w+2) \hat{h}$ \\ \hline
  		$\frac{e^\delta s^\eta}{2\sqrt{\lambda\pi}}  
  		e^{-\frac{1}{4\lambda}(\ln s)^2}$ & $e^{\lambda(w+\eta)^2 + \delta}$\\ \hline 
  		$sh'+s^2 h''$    & $w^2 \hat{h}$\\ \hline
  		$-sh'-3s^2h''-s^3h^{(3)}$ & $w^3\hat{h}$\\ \hline
  	\end{tabular}
  	\label{table2}
  \end{table}
  
  Here, $\lambda$, $\eta$ and $\delta$ are not related to $w$ or $s$, and $h'$, $h''$ and
  $h^{(3)}$ are the first-order, second-order and third-order derivatives of $h$, 
  respectively.
  
  \section{Derivation of Formulas \eqref{eqn:down-put-p0} and \eqref{eqn:floating-put-p0}}
  \label{app:deviation}
  In this appendix, we give detailed derivation of the formulas \eqref{eqn:down-put-p0} and \eqref{eqn:floating-put-p0}.
  
  \subsection{Derivation of formula \eqref{eqn:down-put-p0}}
  
  From Eq. \eqref{eqn:down-p0-mellin}, we know that
  \begin{eqnarray*}
  P_0\left(t, s\right)
  &=&\int_{B}^{K} \left(K-u\right)e^{\delta} \left(\frac{s}{u}\right)^{\eta} 
  \left(\frac{1}{2\sqrt{\lambda\pi}} e^{-\frac{1}{4\lambda}
  \left(\ln \left(\frac{s}{u}\right)\right)^2} \right)\frac{du}{u}-\\[0.5em]
  &&\int_{\frac{B^2}{K}}^{B} \left(\frac{B}{u} \right)^{k_1-1} \left(K-\frac{B^2}{u}
  \right) e^{\delta} \left(\frac{s}{u}\right)^{\eta} \left(\frac{1}{2\sqrt{\lambda\pi}} 
  e^{-\frac{1}{4\lambda}\left(\ln \left(\frac{s}{u}\right)
  \right)^2} \right)\frac{du}{u}. 
  \end{eqnarray*}
  By letting $v=\ln u$, we convert the first integral to
  \begin{eqnarray*}
  &&\int_{\ln B}^{\ln K} (K-e^v) s^{\eta} e^{\delta} e^{-\eta v} 
  \left(\frac{1}{2\sqrt{\lambda\pi}}e^{-\frac{1}{4\lambda}(\ln s-v)^2}\right) dv\\
  &=&\frac{s^{\eta} e^{\delta}}{2\sqrt{\lambda\pi}} \left(\int_{\ln B}^{\ln K} Ke^{-\frac{1}{4\lambda}(v^2-2v\ln s+(\ln s)^2+4\lambda \eta v)} dv\right.\\
  &&\left. -\int_{\ln B}^{\ln K} e^{-\frac{1}{4\lambda}(v^2-2v\ln s+(\ln s)^2+4\lambda 
  (\eta-1)v)} dv\right)\\
  &=&\frac{s^{\eta} e^{\delta}}{2\sqrt{\lambda\pi}} \left(\int_{\ln B}^{\ln K} Ke^{-\frac{1}{4\lambda}(v-\ln s+2\lambda \eta)^2+\lambda \eta^2-\eta \ln s} dv\right.\\
  &&\left.-\int_{\ln B}^{\ln K} e^{-\frac{1}{4\lambda}[v-\ln s+2\lambda (\eta-1)]^2+\lambda (\eta-1)^2-(\eta-1) \ln s} dv\right),
  \end{eqnarray*}
  we further apply the following changes of variables
  \[
  x' := \frac{v-\ln s+2\lambda \eta}{\sqrt{2\lambda}} \quad \mbox{and} \quad 
  x'':=\frac{v-\ln s+2\lambda (\eta-1)}{\sqrt{2\lambda}}
  \]
  to get
  \begin{eqnarray*}
  &&\int_{\ln B}^{\ln K} (K-e^v) s^{\eta} e^{\delta} e^{-\eta v} 
  \left(\frac{1}{2\sqrt{\lambda\pi}}e^{-\frac{1}{4\lambda}(\ln s-v)^2}\right) dv\\
  &=& \frac{e^\delta}{\sqrt{2\pi}} \left(Ke^{\lambda \eta^2}\int_{\frac{\ln(\frac{B}{s})+
  2\lambda \eta}{\sqrt{2\lambda}}}^{\frac{\ln(\frac{K}{s})+2\lambda \eta}{\sqrt{2\lambda}}} 
  e^{-\frac{x'^2}{2}}dx'-se^{\lambda(\eta-1)^2}\int_{\frac{\ln(\frac{B}{s})+
  2\lambda(\eta-1)}{\sqrt{2\lambda}}}^{\frac{\ln(\frac{K}{s})+2\lambda (\eta-1)}{\sqrt{2\lambda}}} e^{-\frac{x''^2}{2}}dx''\right)\\
  &=&Ke^{\delta+\lambda\eta^2}\left[\Phi\left(\frac{\ln(\frac{K}{s})+2\lambda \eta}{\sqrt{2\lambda}}\right)-\Phi\left(\frac{\ln(\frac{B}{s})+2\lambda \eta}{\sqrt{2\lambda}}\right)\right]\\
  &&-se^{\delta+\lambda(\eta-1)^2} \left[\Phi\left(\frac{\ln(\frac{K}{s})+2\lambda (\eta-1)}{\sqrt{2\lambda}}\right)
  -\Phi\left(\frac{\ln(\frac{B}{s})+2\lambda (\eta-1)}{\sqrt{2\lambda}}\right)\right].\\
  \end{eqnarray*}
  Now, if we plug into $\delta$, $\eta$ and $\lambda$ into the above formula, we derive 
  \begin{eqnarray*}
  &&\int_{\ln B}^{\ln K} (K-e^v) s^{\eta} e^{\delta} e^{-\eta v} 
  \left(\frac{1}{2\sqrt{\lambda\pi}} e^{-\frac{1}{4\lambda}(\ln s-v)^2}\right) dv\\
  &&=K e^{-r\left(T-t\right)} \left[\Phi\left(-\Delta_{-}\left(\frac{s}{K}\right)\right)-\Phi\left(-\Delta_{-}
  \left(\frac{s}{B}\right)\right)\right]\\
  &&-s\left[\Phi\left(-\Delta_{+}\left(\frac{s}{K}\right)\right)-\Phi\left(-\Delta_{+}
  \left(\frac{s}{B}\right)\right) \right].
  \end{eqnarray*}
  Similarly, we can evaluate the second integral
  \[
  \int_{\frac{B^2}{K}}^{B} \left(\frac{B}{u} \right)^{k_1-1} \left(K-\frac{B^2}{u}\right) 
  e^{\delta} \left(\frac{s}{u}\right)^{\eta} \left(\frac{1}{2\sqrt{\lambda\pi}} e^{-\frac{1}{4\lambda}\left(\ln \left(\frac{s}{u}\right)\right)^2} \right)\frac{du}{u}
  \]
  to obtain
  \begin{eqnarray*}
  && K e^{-r\left(T-t\right)}\left(\frac{B}{s}\right)^{k_1-1} \left[\Phi\left(\Delta_{-}\left(\frac{B}{s}\right)\right)-\Phi\left(\Delta_{-}
  \left(\frac{B^2}{sK}\right)\right)\right]\\
  &&-B\left(\frac{B}{s}\right)^{k_1} \left[\Phi\left(\Delta_{+}\left(\frac{B}{s}\right)\right)-\Phi\left(\Delta_{+}
  \left(\frac{B^2}{sK}\right)\right) \right].
  \end{eqnarray*}
  
  Putting these two integrals together yields formula \eqref{eqn:down-put-p0}. 
  
  \subsection{Derivation of formulas \eqref{eqn:floating-put-p0}}
  
  From Eq. \eqref{eqn:floating-q0-mellin}, we have
  
  \begin{eqnarray*} 
  Q_0(t, u) 
  &=&\int_{0}^{1} \left(1-\xi\right)e^{\delta} \left(\frac{u}{\xi}\right)^{\eta} 
  \left(\frac{1}{2\sqrt{\lambda\pi}} e^{-\frac{1}{4\lambda}
  	\left(\ln \left(\frac{u}{\xi}\right)\right)^2} \right)\frac{d\xi}{\xi} + \\[0.5em]
  &&\int_{1}^{\infty} \left(-\frac{1}{k_1}\xi^{1-k_1} +\frac{\xi}{k_1} \right)e^{\delta} 
  \left(\frac{u}{\xi}\right)^{\eta} \left(\frac{1}{2\sqrt{\lambda\pi}}
  e^{-\frac{1}{4\lambda}\left(\ln \left(\frac{u}{\xi}\right)\right)^2} \right)
  \frac{d\xi}{\xi}. \nonumber
  \end{eqnarray*}
  We let $v=\ln \xi$. For the first integral, we have
  \begin{eqnarray*}
  &&\int_{0}^{1} (1-\xi)e^{\delta} \left(\frac{u}{\xi}\right)^{\eta} \left(\frac{1}{2\sqrt{\lambda\pi}} e^{-\frac{1}{4\lambda}\left(\ln \left(\frac{u}{\xi}\right)\right)^2} \right)\frac{du}{u}\\
  &=&\int_{-\infty}^{0} u^{\eta}\left(1-e^v\right) e^{\delta -v\eta} \left(\frac{1}{2\sqrt{\lambda\pi}} e^{-\frac{1}{4\lambda}(\ln u-v)^2} \right) dv\\
  &=& \frac{u^{\eta} e^{\delta}}{2\sqrt{\lambda\pi}} \left(\int_{-\infty}^{0} e^{-\frac{1}{4\lambda}\left(v^2-2v\ln u+\left(\ln u\right)^2+4\lambda \eta v\right)} dv \right.\\
  &&\left.-\int_{-\infty}^{0} e^{-\frac{1}{4\lambda}\left(v^2-2v\ln u+\left(\ln u\right)^2
  +4\lambda \left(\eta-1\right) v\right)} dv\right)\\
  &=&\frac{u^{\eta} e^{\delta}}{2\sqrt{\lambda\pi}}\left(\int_{-\infty}^{0} e^{-\frac{1}{4\lambda}\left(v-\ln u+2\lambda \eta\right)^2+\lambda \eta^2-\eta \ln u} dv\right.\\
  &&\left.-\int_{-\infty}^{0} e^{-\frac{1}{4\lambda}\left(v-\ln u +2\lambda \left(\eta-1\right)\right)^2+\lambda \left(\eta-1\right)^2-\left(\eta-1\right) \ln u} dv\right).
  \end{eqnarray*}
  Next, we let
  \[
  v':= \frac{v-\ln u +2\lambda \eta}{\sqrt{2\lambda}}\quad \mbox{and} \quad 
  v'':=\frac{v-\ln u+2\lambda \left(\eta-1\right)}{\sqrt{2\lambda}}.
  \]
  Then, we have
  \begin{eqnarray*}
  &&\int_{0}^{1} (1-\xi)e^{\delta} \left(\frac{u}{\xi}\right)^{\eta} \left(\frac{1}{2\sqrt{\lambda\pi}} e^{-\frac{1}{4\lambda}\left(\ln \left(\frac{u}{\xi}\right)\right)^2} \right)\frac{du}{u}\\
  &=& \frac{e^{\delta}}{\sqrt{2\pi}} \left(\int_{-\infty}^{\frac{-\ln u+2\lambda \eta}{\sqrt{2\lambda}}} e^{-\frac{v'^2}{2}+\lambda \eta^2} dv'
  -u\int_{-\infty}^{\frac{-\ln u+2\lambda (\eta-1)}{\sqrt{2\lambda}}} e^{-\frac{v''^2}{2}+\lambda \left(\eta-1\right)^2} dv''\right)\\
  &=& e^{\delta+\lambda \eta^2}\Phi\left(\frac{-\ln u+2\lambda \eta}{\sqrt{2\lambda}}\right)
  -ue^{\delta + \lambda \left(\eta-1\right)^2} \Phi\left(\frac{-\ln u +2\lambda \left(\eta-1\right)}{\sqrt{2\lambda}}\right) \\
  &=&e^{-r\left(T-t\right)} \Phi\left(-\Delta_{-}\left(\frac{s}{z}\right)\right)-
  \left(\frac{s}{z}\right) \Phi\left(-\Delta_{+}\left(\frac{s}{z}\right)\right).
  \end{eqnarray*}

  For the second integral, we have
  \begin{eqnarray*}
  &&\int_{1}^{\infty} \left(-\frac{1}{k_1} \xi^{1-k_1} +\frac{\xi}{k_1} \right)e^{\delta} \left(\frac{u}{\xi}\right)^{\eta} \left(\frac{1}{2\sqrt{\lambda\pi}} e^{-\frac{1}{4\lambda}\left(\ln \left(\frac{u}{\xi}\right)\right)^2} \right)\frac{d\xi}{\xi}\\
  &=&\int_{0}^{\infty} \left(-\frac{1}{k_1} e^{(1-k_1)v}+\frac{1}{k_1} e^{v}\right)e^{\delta} u^{\eta} e^{-v\eta}\left(\frac{1}{2\sqrt{\lambda\pi}} e^{-\frac{1}{4\lambda}
  \left(\ln u-v\right)^2}\right) dv\\
  &=& \frac{e^{\delta} u^{\eta}}{2k_1\sqrt{\lambda\pi}} 
  \int_{0}^{\infty} \left(-e^{\eta v -\frac{1}{4\lambda}\left(\ln u-v\right)^2}+e^{v\left(1-\eta\right)-\frac{1}{4\lambda}\left(\ln u-v\right)^2} \right)dv\\
  &=& \frac{e^{\delta} u^{\eta}}{2k_1\sqrt{\lambda\pi}} 
  \left(\int_{0}^{\infty} -e^{-\frac{1}{4 \lambda}\left(v-\ln u-2\lambda\eta\right)^2
  +\lambda\eta^2 + \eta\ln u} dv\right.\\
  &&\left.+\int_{0}^{\infty} e^{-\frac{1}{4 \lambda}\left(v-\ln u -2\lambda (1-\eta)\right)^2 +\lambda (1-\eta)^2+ (1-\eta)\ln u} dv\right),\\
  \end{eqnarray*}
  where we use the fact that $k_1-1 +\eta =-\eta$. Further, we introduce a new variable
  \[
  v''':=\frac{v-\ln u-2\lambda\eta}{\sqrt{2\lambda}}.
  \]
  Then, we have
  \begin{eqnarray*}
  &&\int_{1}^{\infty} \left(-\frac{1}{k_1} \xi^{1-k_1} +\frac{\xi}{k_1} \right)e^{\delta} \left(\frac{u}{\xi}\right)^{\eta} \left(\frac{1}{2\sqrt{\lambda\pi}} e^{-\frac{1}{4\lambda}\left(\ln \left(\frac{u}{\xi}\right)\right)^2} \right)\frac{d\xi}{\xi}\\
  &=& \frac{e^{\delta} u^{\eta}}{k_1\sqrt{2\pi}} 
  \left(\int_{\frac{-\ln u-2\lambda\eta}{\sqrt{2\lambda}}}^{\infty} -e^{-\frac{v'''^2}{2}} e^{\lambda\eta^2 +\eta\ln u} dv'''\right.\\
  &&\left.+\int_{\frac{-\ln u+2\lambda\left(\eta-1\right)}{\sqrt{2\lambda}}}^{\infty} 
  e^{-\frac{v''^2}{2}} e^{\lambda\left(\eta-1\right)^2+\left(1-\eta\right)\ln u} dv''\right)\\
  &=&-\frac{1}{k_1} e^{\delta+\lambda\eta^2} u^{1-k_1}
  \Phi\left(\frac{\ln u+2\lambda\eta}{\sqrt{2\lambda}}\right)
  +\frac{1}{k_1} u e^{\delta+\lambda\left(\eta-1\right)^2} \Phi\left(\frac{\ln u +2\lambda\left(1-\eta\right)}{\sqrt{2\lambda}}\right) \\
  &=&-\frac{1}{k_1} \left(\frac{s}{z}\right)^{1-k_1} e^{-r\left(T-t\right)} \Phi\left(-\Delta_{-}\left(\frac{z}{s}\right)\right)
  +\frac{1}{k_1} \left(\frac{s}{z}\right) \Phi\left(\Delta_{+}\left(\frac{s}{z}\right)\right).
  \end{eqnarray*}
  Putting these two integrals together and using the fact that $P_0 = z Q_0$, we can obtain  
  our formula \eqref{eqn:floating-put-p0}. 
  
  \bigskip
  \centerline{\sc Acknowledgment}
  
  \medskip
  The author J.-H. Kim gratefully acknowledges the financial support by the National 
  Research Foundation of Korea grant NRF2021R1A2C10040.

  \renewcommand\bibname{references}
  \singlespacing
  \bibliographystyle{apacite}
  \bibliography{reference}

  \end{document}